\documentclass{amsart}
\usepackage{fullpage}
\usepackage{amssymb,amsmath}
\usepackage{amsthm}
\usepackage{epsfig,graphicx}
\usepackage{mathtools}
\usepackage{color}
\usepackage{cite}
\usepackage{array}
\usepackage[version=4]{mhchem}
\usepackage[numbers,sort&compress]{natbib} 
\usepackage{setspace}

\newcommand{\Cov}{\mathrm{Cov}}
\newcommand{\bA}{\mathbf A}
\newcommand{\bAst}{\langle \mathbf A\rangle_{\bast}}
\newcommand{\ba}{\boldsymbol \alpha}
\newcommand{\bast}{\bast}

\newcommand*{\bbbe}{\mathbb{E}}
\newcommand*{\dint}{\mathrm{d}}

\newcommand{\md}{d\kern-0.035cm\char39\kern-0.03cm}

\def\ba{\ensuremath\boldsymbol\alpha}
\def\bast{\ensuremath\boldsymbol\alpha^\ast}

\begin{document}

\title{Dynamic Maximum Entropy provides accurate approximation of structured population dynamics}
\maketitle

\begin{center}
{Katar{\'\i}na Bo{\md}ov\'a$^{a\ast}$, Enik\H{o} Sz\'ep$^b$, Nicholas H. Barton$^b$}
\vskip 0.3cm
$^a${Faculty of Mathematics, Physics and Informatics, Comenius University, \\Mlynsk\'a Dolina, 84248 Bratislava, Slovakia},\\
$^b${Institute of Science and Technology Austria (IST Austria), \\Am Campus 1, Klosterneuburg A-3400, Austria}\\
{\it $^\ast$katarina.bodova@fmph.uniba.sk}
\end{center}

\begin{abstract}
Realistic models of biological processes typically involve interacting components on multiple scales, driven by changing environment and inherent stochasticity. Such models are often analytically and numerically intractable. We revisit a dynamic maximum entropy method that combines a static maximum entropy and a quasi-stationary approximation. This allows us to reduce stochastic non-equilibrium dynamics expressed by the Fokker-Planck equation to a simpler low-dimensional deterministic dynamics, without the need to track microscopic details. Although the method has been previously applied to a few (rather complicated) applications in population genetics, our main goal here is to explain and to better understand how the method works. We demonstrate the usefulness of the method for two widely studied stochastic problems, highlighting its accuracy in capturing important macroscopic quantities even in rapidly changing non-stationary conditions. For the Ornstein-Uhlenbeck process, the method recovers the exact dynamics whilst for a stochastic island model with migration from other habitats, the approximation retains high macroscopic accuracy under a wide range of scenarios for a dynamic environment.

\end{abstract}
\section{Introduction}

Conceptual understanding of realistic problems in applied sciences is often hindered by the curse of complexity, with quantities of interest coupling to finer features. Due to their multiscale character even simple questions lead to exploration of the full complexity of the system. But how can we ever understand the processes around us if incremental learning is impossible?

\begin{figure}[h]
	\centering
	\includegraphics[width=\textwidth]{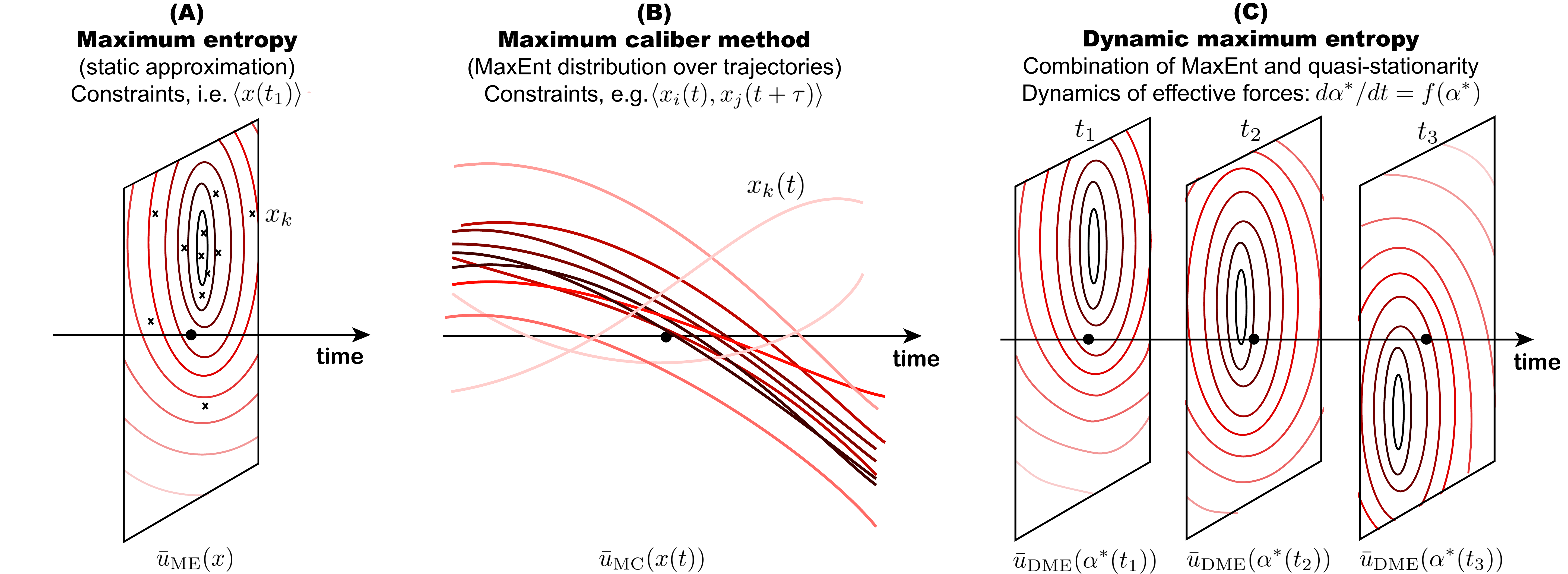}
	\caption{Variational methods ME and MC compared to DME. 
	(A) ME looks at a snapshot $x$ of a process at a particular time and provides an approximation $\bar u_\text{ME}(\mathbf x)$ of 
	the microscopic distribution, given knowledge of a few key macroscopic observables.
	(B) MC is analogous to ME, however, each data point represents a trajectory $x(t)$. MC connects 
	the microscopic distribution over possible trajectories with macroscopic constraints and approximates it by $\bar u_\text{MC}(x(t))$.
	(C) DME is a quasi-stationary approximation of the stochastic dynamics, given by the FPE, which
	reduces the full problem to a low-dimensional dynamics. 
	This reduction is a consequence of a ME ansatz; the approximation at each time $\bar u_\text{DME}(\ba(t))$ 
	solves the ME problem (stationary form in the FPE), 
	where the dynamics of the effective forces $\ba$ are systematically derived from the FPE.
	\label{fig:methods}}
\end{figure}

Statistical mechanics provides a clever way to understand complex multiscale problems by linking processes on different scales through the parsimony principle -- the method of maximum entropy (ME), introduced by \cite{Jaynes1957}. ME has the form of a variational problem  where an entropy of the microscopic distribution is maximized, while enforcing macroscopic constraints, e.g. average energy of gas particles \citep{Jaynes1957}, see figure~\ref{fig:methods}A. 
The method gained popularity in applied sciences in recent decades primarily as a tool for inference from empirical data, for instance in bird flocking \citep{bialek2012}, neuronal firing \citep{schneidman2006}, or protein variability \citep{mora2010}.

However, realistic biological questions often do not adhere to the assumption of stationarity. Adaptation of populations to spatial
 and temporal ecological gradients is an example of a complex non-equilibrium processes in which ecological and evolutionary processes interact \cite{thompson98,schoener2011,uecker2018,sachdeva2019,polechova2018}. 
How can a simple concept of ME be applied to such systems? The most straightforward approach is to use ME 
 on dynamic trajectories, forcing constraints on the dynamical features (figure~\ref{fig:methods}B). This approach, called maximum caliber (MC), introduced by \cite{Jaynes1980} is suitable for inference of models and their parameters from dynamic data. Comprehensive reviews of MC \cite{Presse2013,Dixit2018,Ghosh2020} provide multiple examples where the method has been successfully applied to non-stationary biological processes. 
 
While such an inverse approach is useful for understanding temporal data, in many cases data are not available but the dynamics, although often extremely complicated, are known to be accurately described by the Fokker-Planck equation (FPE). The aim of our work is to demonstrate usefulness of a theoretical dimensionality-reduction technique, which approximates possibly many-dimensional stochastic dynamics to a low-dimensional deterministic dynamics of a few key observable quantities (figure~\ref{fig:methods}C). 
The approximation combines static ME ansatz in the FPE equation with a quasi-stationary assumption. We refer to this method as the dynamic ME (DME). DME has been used before to solve problems in quantitative genetics \cite{barton2009,vladar2011stab,Bodova2016} and independently in cosmology \cite{Hick1987,Tsallis1996,Plastino1997b,Plastino1997a}. 

The most surprising feature of DME is its accuracy. The method, derived from the assumption of quasi-stationarity, remains extremely accurate even in far-from-equilibrium regimes \citep{barton2009,vladar2011stab,Bodova2016,Bodova2018}. Nevertheless, explicit estimates of the method's accuracy opens a challenging mathematical problem. While in reaction kinetics, where the quasi-stationary assumption is often used to reduce dynamics of reactants by assuming that concentration of certain chemicals does not change at the considered timescale, its validity has been shown rigorously only for a few basic systems using singular perturbation \citep{SS1989,walcher2013,kollar2015}. In comparison, the accuracy of DME still remains a mystery. We will address this puzzle by focusing on two processes: the Ornstein-Uhlenbeck (OU) process, for which DME gives an exact solution, and a logistic model of population growth in a continent-island model, which is one of the simplest population models that includes geographic population structure. These processes are explored in the far-from equilibrium regime.

\section{Dynamic maximum entropy}
\noindent 
Here we present a dynamic maximum entropy (DME) method to approximate stochastic dynamics in terms of a FPE \citep{Hick1987,Tsallis1996,Plastino1997b,Plastino1997a,barton2009,vladar2011stab,Bodova2016,Bodova2018}. The method is based on a combination of ME in statistical physics  \citep{Jaynes1957}, which solves the stationary problem exactly, with a quasi-stationary assumption, as typically used in chemical kinetics \citep{SS1989} to reduce the number of equations. The method applies to stochastic dynamics with an explicit stationary distribution, even though its application is not limited to such problems (as shown in \cite{Bodova2018} a solution ansatz, which is not based on the stationary form can sometimes lead to more accurate approximation). DME was introduced in population genetics to understand how quantitative traits change in time in the presence of various evolutionary mechanisms without resolving details about the dynamics of the underlying gene frequencies. Independent use of the method in statistical physics focused on exact and approximate solutions of the nonlinear FPE arising in cosmology. We provide a comprehensive summary of the method based on \cite{Bodova2018} in this section. 

Assume stochastic dynamics in the Langevin form
	\begin{equation}
		\mathrm{d} x_k(t) =   \frac{g^2(x_k)}{2}   
		\frac{\partial }{\partial x_k}  
		\left [ \sum_{i=1}^d  \alpha_i A_i(\mathbf x) \right ]
		\mathrm{d} t + g(x_k) \mathrm{d} \xi_k(t)\,. 
		\label{eq:Langevin}
	\end{equation}
for $\mathbf x = (x_1,\dots, x_N)$, $ x_k\in \Omega_X$, $t>0$, and $\mathbf x(0) = \mathbf x_0$.  The potential in the first term is a linear combination of forces $\alpha_i$, acting on functions $A_i(\mathbf x)$, which may introduce coupling between equations. Function $g(x_k)$ represents amplitude of stochastic fluctuations and $\xi_k(t)$ are independent Wiener processes. 
Previous studies (e.g. \cite{barton2009,Vladar2011a,vladar2011stab,Bodova2016}) focused on examples in population genetics where $x_k$ corresponds to the frequency of a certain gene, affecting some quantitative trait. This frequency depends on evolutionary processes, e.g. selection, mutation, and inherent stochastic fluctuations, described by the forces $\alpha_i$.
At this point we consider constant forces $\alpha_i \in \mathbb{R}$, which means that the distribution $u(t,\mathbf x)$ follows dynamics described by the FPE
	\begin{equation} 
		\frac{\partial u(t,\mathbf x)}{\partial t} = 
		- \sum_{k=1}^N \sum_{i=1}^d  \alpha_i  \frac{\partial}{\partial x_k} \left[ \frac{g^2(x_k)}{2}\,
		\frac{\partial A_i(\mathbf x)}{\partial x_k} \,u(t,\mathbf x) \right]
		+ \frac12 \sum_{k=1}^N \frac{\partial^2}{\partial x_k^2} \left[g^2(x_k) u(t,\mathbf x) \right]\,, 
		\label{eq:FPE}
	\end{equation}
which can be also expressed in the flux form $\partial_t  u(t,\mathbf x) = -\sum_{k=1}^N \partial_{x_k} J_k[t,\mathbf x]$. This FPE is complemented with no-flux boundary conditions $J_k[t,\mathbf x] = 0$ at $ x_k \in \partial \Omega_X$ and the initial condition $u(0,\mathbf x) = u_0(\mathbf x)$.

\subsection{Stationary solution and ME} At large times the dynamics approach a stationary distribution, parametrized by the vector of forces $\ba = (\alpha_1,\dots,\alpha_d) \in \mathbb R^d$
	\begin{equation} 
		 \bar u_{\ba}(x) = \frac{1}{\mathcal{Z}_{\ba}} \left( \prod_{k=1}^N \frac{1}{g^2(x_k)} \right)
		 \exp \left[ \sum_{i=1}^d  \alpha_i A_i(\mathbf x) \right] 
		 \label{eq:equilibrium}
	\end{equation}
with the normalization coefficient (i.e., the partition function)
	\begin{equation} 
		 \mathcal{Z}_{\ba} = \int_{\Omega_X^N} \left( \prod_{k=1}^N \frac{1}{g^2(x_k)} \right)
		 \exp \left[ \sum_{i=1}^d  \alpha_i A_i(\mathbf x) \right] \mathrm{d}\mathbf x\,, 
		 \label{eq:Z}
	\end{equation}
where $\mathbf{A} = (A_1,\dots,A_d)$ 
is a vector function of the state variables $\mathbf x$ which the forces $\alpha_i$ acts on. Using the terminology in statistical physics we refer to functions $A_i$ as observables, as their expectations  in problems in physics provide a macroscopic description of the system in terms of its natural observable quantities (i.e., average energy of a gas particle, as formulated by \cite{Jaynes1957}). We extend our scope from looking at a problem with constant forces $\ba$ to time-dependent forces $\ba(t)$ to account for realistic scenarios where the dynamics, initially settled to a stationary solution, are pushed out-of equilibrium by changes in the forces $\ba$. 
The dynamics of the expectation $\langle A_j \rangle$ follows from the FPE.
Using notation 
$B_{ji} = \left  \langle \sum_{k=1}^N \frac{g(x_k)}{2}  \frac{\partial A_j}{\partial x_k}  \frac{\partial A_i}{\partial x_k}  \right \rangle$, $V_j =  \left  \langle  \sum_{k=1}^N g^2(x_k)  \frac{\partial^2 A_j}{\partial x_k^2}  \right \rangle$ 
we obtain 
	\begin{eqnarray}
		\frac{\partial}{\partial t} \langle A_j \rangle  &=& 
		\sum_{i=1}^d   B_{ji} \alpha_i + \frac{1}{2} V_j \,.
		\label{eq:Amean_simp}
	\end{eqnarray}
This forms a system of ordinary differential equations for $\langle \mathbf A \rangle$, which is generally not closed due to nonlinearity of the functions $A_j(\mathbf x)$. 
Next we define a logarithmic relative entropy
	\begin{equation} 
		H[u|\bar u_{\ba}] := \int_{\Omega^N_X} u \ln \frac{u}{\bar u_{\ba}} \mathrm{d}\mathbf x\,,
		\label{eq:rel_entropy}
	\end{equation}
where $u(t,\mathbf x)$ is a solution of the FPE at time $t \ge 0$ and $x_k\in \Omega_X$. 
For any $t \ge 0$ relative entropy \eqref{eq:rel_entropy} has a maximum at $\ba = \ba^\ast$, which can be obtained by solving a set of first-order conditions
	\begin{equation} 
		0 = \frac{d}{d\alpha_i} H[u|\bar u_{\ba}] =
	  -\int_{\Omega^N_X}  \frac{u}{\bar u_{\ba}} \frac{d}{d\alpha_i} \bar u_{\ba} \mathrm{d}\mathbf x 
		= \langle A_i \rangle_{\bar u_{\ba}} -  \langle A_i \rangle_{u}\,.
		\label{eq:criticality}
	\end{equation}
The above intriguing relationship states that if for a given time $t$ there exists a maximum of the relative entropy \eqref{eq:rel_entropy} with respect to all $\alpha_i$ reached for some choice of parameters $\ba^\ast$, then the expectation of $A_i$ through the distribution $u(t,\mathbf x)$  equals the expectation through the stationary distribution $\bar u_{\ba^\ast}(\mathbf x)$ at this time. This simple fact, shown previously in \cite{Bodova2018} and in a slightly different form by \cite{barton2009,vladar2011stab,Bodova2016} suggests that instead of the full representation of the problem using FPE one could trace only the $d$-dimensional dynamics of $\ba^\ast$, that parametrize the approximate solution by the form \eqref{eq:equilibrium}. Furthermore, the two representations agree in terms of the expectations $\langle A_i\rangle$ at a given time. Note that the solvability of the equation \eqref{eq:criticality} (for $\bast$) is nontrivial \citep{Bodova2018} and requires further attention. Nevertheless, concavity of the relative entropy is implied by the following relationship
	\begin{equation}
		\frac{d^2}{d\alpha_i d\alpha_j} H[u|\bar u_{\ba}] 
		=   \int_{\Omega^N_X} \left[  \frac{d}{d\alpha_j}\ln  \bar u_{\ba}\right ] 
		\left [  \frac{d}{d\alpha_i} \ln \bar u_{\ba} \right ]  u \mathrm{d} \mathbf x
		 -\int_{\Omega^N_X} \left [ \frac{1}{\bar u_{\ba}} \frac{d^2}{d\alpha_i d\alpha_j} \bar u_{\ba}  \right ] u \mathrm{d} \mathbf x 
		 = \Cov(A_i,A_j)_{\bar u_{\alpha}}
		\label{eq:hessian}
	\end{equation}
analogous to similar expressions in \cite{barton2009,vladar2011stab,Bodova2016,Bodova2018} stating that the Hessian of the relative entropy is positive semidefinite.

\subsection{Dynamical approximation} We have established a relationship between the solution of the full stochastic dynamics \eqref{eq:FPE} and a stationary form $u_{\ba^\ast}$ parametrized by suitable effective forces $\ba^\ast$ following ME.
However, as we demonstrated in figure~\ref{fig:methods}A, ME is applicable only to static problems. 
When the system is out-of-equilibrium, we need to establish a dynamic relationship between the values $\ba^\ast(t_1)$ and $\ba^\ast(t_2)$ for $t_1 \neq t_2$ by using the information captured by the FPE. 

To derive the DME approximation of \eqref{eq:FPE} we use an ansatz $u(t,\mathbf x) = \bar u_{\ba(t)}(\mathbf x) + R(t,\mathbf x)$ for some continuous $\ba(t)$ where $R(t,\mathbf x)$ is the time-dependent residual. The dynamics of the expectations \eqref{eq:Amean_simp} becomes 
	\begin{equation}
		\frac{\partial}{\partial t} \langle \mathbf A \rangle_{\ba}  = 		
		 \mathbf B_{\ba}   \ba(t) + \frac{1}{2} \mathbf V_{\ba}
		 +\left[ \mathbf B_R \ba (t) + \frac{1}{2} \mathbf V_R
		  - \frac{\partial}{\partial t} \langle \mathbf A \rangle_R \right]\,.
		\label{eq:Amean1}
	\end{equation}
where $\langle\cdot \rangle_u$ represents expectation through distribution $u$ and $\langle \cdot \rangle_{\ba} := \langle \cdot \rangle_{\bar u_{\ba(t)}}$.
Now we make two key assumptions. First, we assume that the residual terms in the bracket of \eqref{eq:Amean2} are small and we neglect them. In addition, we also impose a quasi-stationarity approximation, assuming that $\ba^\ast$ are chosen to satisfy the equilibrium relationship 
	\begin{equation}
		 \mathbf B_{\bast}   \ba^\ast(t) + \frac{1}{2} \mathbf V_{\bast} = 0
		\label{eq:QS}
	\end{equation}  
for all $t>0$. Both steps are easy to justify if the forces $\ba(t)$ are slowly changing (in the adiabatic regime) and the solution of the FPE is thus close to an equilibrium form \eqref{eq:equilibrium}. However, its validity when out-of-equilibrium is not clear. We use \eqref{eq:QS} to replace $ \mathbf V_{\ba}$ by $\mathbf V_{\bast}$ and $\mathbf B_{\ba}$ by $\mathbf B_{\bast}$ in \eqref{eq:Amean1} to approximate 
	\begin{equation}
		\frac{\partial}{\partial t} \langle \mathbf A \rangle_{\bast}  \approx 		
		 \mathbf B_{\bast}   (\ba(t)-\bast) \,.
		\label{eq:Amean2}
	\end{equation}
Finally, to obtain a closed dynamical system for $\ba^\ast$ we use the chain rule, noting that 
$\frac{\partial \langle \mathbf A \rangle_{\bast}}{\partial t}  = \frac{\partial \langle \mathbf A \rangle_{\bast}}{\partial \bast} \frac{\partial \bast}{\partial t}$. Combined equations \eqref{eq:criticality} and \eqref{eq:hessian} imply that differentiation of an expectation $\langle A_i \rangle_{\bast}$ with respect to $\alpha_j$ gives us a covariance $\mathbf C_{\bast} := \Cov(A_i,A_j)_{\bast}$. Therefore
	\begin{equation}
		\frac{\partial \bast}{\partial t} =
		\mathbf C_{\bast}^{-1}  \mathbf B_{\bast}   (\ba(t)-\bast) \,, \qquad \ba(0) = \ba_0
		\label{eq:DME}
	\end{equation}
together with a parametric form \eqref{eq:equilibrium} represents DME approximation of dynamics \eqref{eq:FPE}. Solution of \eqref{eq:DME} can be plugged into the stationary parametric form \eqref{eq:equilibrium}, which allows us to not only study the accuracy of the key moments used in DME, but also to compute any statistical feature of the approximate solution and compare it with the exact solution. 

Note that the equation \eqref{eq:DME} can be solved for any prescribed continuous function $\ba(t)$ and in the most extreme case, the forces $\ba(t)$ can contain step changes, which clearly violate the quasi-stationarity assumption. However, all previously studied applications of DME approach showed that the approximation captures extremely well the expectations of the key functions even when the forces change rapidly. This is one of the most remarkable and unexpected features of the DME approach which will be studied here. 

Realistic situations (e.g. selection acting on quantitative traits that depend on many alleles) typically involve high-dimensional stochastic dynamics with nonlinearities and coupling terms \cite{barton2009,vladar2011stab,Bodova2016}. However, for simplicity we consider examples leading to a simpler one-dimensional form with a remark that more complex models can be analyzed using this approach as long as the stationary distribution of the FPE is explicit. However, even when $\mathbf x$ is a scalar $\bA$ and $\ba$ are vectors in all problems studied here: these vectors summarise the infinite-dimensional distribution of $\mathbf x$. 

\section{Ornstein-Uhlenbeck process}

\subsection{The model} Here we outline a simple example of stochastic dynamics where DME reproduces the exact dynamics. We use an OU process, which describes dynamics with linear relaxation to an equilibrium in the presence of constant Gaussian noise (examples include a particle under friction, animal motion, financial time series, etc.). The OU process has three parameters: $\mu$ -- long-time average of the state variable $x$, $\beta$ -- persistence length, and $\sigma$ -- magnitude of noise. It has the form
\begin{equation}
	dx = \beta(\mu-x)dt + \sigma d\xi(t)\,.
	\label{eq:OU}
\end{equation}
The stationary distribution of the equation \eqref{eq:OU} can be obtained from the FPE, which describes time-evolution of the probability distribution of $x$, denoted by $u(t,x)$
\begin{equation}
	\frac{\partial}{\partial t} u = -\frac{\partial}{\partial x} \left[ \beta(\mu-x)u \right]+\frac{\sigma^2}{2} \frac{\partial^2}{\partial x^2} u\,,
	\label{eq:OU_FPE}
\end{equation}
by setting the left-hand side equal to 0
\begin{equation}
	\bar u_{\ba}(x)= \frac{1}{\mathcal Z} \exp \left[\frac{2\mu\beta}{\sigma^2}x -  \frac{\beta}{\sigma^2}x^2 \right] 
	= \frac{1}{\mathcal Z} \exp \left[ \frac{1}{\sigma^2} \boldsymbol \alpha \cdot \mathbf A \right]\,,
	\label{OU_stationary}
\end{equation}
where $\mathcal Z = 1/\sqrt{\beta/(2\pi\sigma^2)}$ is the normalization factor, $\ba =  (2\mu \beta,-\beta)$ and $\mathbf A = (x,x^2)$.
If we set the initial condition of \eqref{eq:OU} as the stationary distribution corresponding to the parameters $(\beta_0, \mu_0, \sigma_0)$,  i.e., as a Gaussian $x_0 \sim \mathcal{N}(\mu_0,\sigma_0^2/2\beta_0)$, then the solution of \eqref{eq:OU_FPE} is a Gaussian $\mathcal{N}(m(t),v(t))$ for every $t\ge 0$. The functions $m(t)$ and $v(t)$ represent a time-dependent mean and variance, which solve the system of ODEs
\begin{align}
	\dot m&= \beta(\mu-m)\,,  \qquad 
	\dot v=\sigma^2 - 2\beta v\,, \label{mvODE}
\end{align}
with initial conditions $m(0) = \mu_0$ and $v(0) = \frac{\sigma^2}{2\beta_0}$.
The explicit solution
\begin{align}
	m(t) &= \mu + (\mu_0-\mu)e^{-\beta t} \,, \label{OUmean}\\
	v(t) &=  \frac{\sigma^2}{2\beta_0} e^{-2\beta t} + \frac{\sigma^2}{2\beta} (1-e^{-2\beta t})\, \label{OUvar}
\end{align}
satisfies $m(t) \rightarrow \mu$ and $v(t) \rightarrow \sigma^2/2\beta$ as $t\rightarrow \infty$. 

The stationary solution $\bar u_{\ba}(x)$ in \eqref{OU_stationary} solves a variational ME problem with the relative entropy defined in \eqref{eq:rel_entropy} using forces $\ba$ and observables $\mathbf A$. Even though the OU process has three natural parameters $(\beta, \mu, \sigma)$ ME implies that the stationary solution is a 2-parameter family of functions of form \eqref{OU_stationary}. 
Equivalently, we may restate the initial distribution by taking a fixed initial value $\sigma_0 = \sigma$ and picking $\beta_0$ such that the initial variance of the Gaussian $\sigma^2/2\beta_0 = v_0$ has the desired initial value. This allows us to keep the volatility of the process fixed in the DME approach. Following the DME derivation steps (details in Appendix~\ref{ap:OU}) we obtain a 2-dimensional dynamical system for the effective forces $\bast$
 shown in \eqref{DMEfull1}-\eqref{DMEfull2}. This coupled dynamical system can be transformed into decoupled dynamics of $\mu^\ast$ and $\beta^\ast$ of the form
\begin{eqnarray}
	\frac{\mathrm{d}\mu^\ast}{\mathrm{d}t}&=& \beta (\mu-\mu^\ast) \label{DME1}\,,\\
	\frac{\mathrm{d}\beta^\ast}{\mathrm{d}t}&=&  2\beta^\ast (\beta - \beta^\ast)\,.	\label{DME2}
\end{eqnarray}
First, note that the dynamical system  \eqref{DME1}-\eqref{DME2}
is independent of the noise magnitude $\sigma$, which is considered constant. It is explicitly solvable since it is decoupled and while the first equation is linear, the second one is logistic. The DME solution 
is consistent with \eqref{OUmean}-\eqref{OUvar} given $\sigma=\sigma_0$, $m(t) = \mu^\ast(t)$ and $v(t) = \sigma^2/(2\beta^\ast(t))$ are functions of effective forces $\bast$. This is due to the linearity of the OU process, which yields a closed dynamics of the first two moments and thus preserves a Gaussian form of the solution at each time, provided we started with a Gaussian initial condition. Note that the OU process is not the only stochastic process where DME provides an exact solution. As \cite{Tsallis1996} showed, the nonlinear extension of the OU process can be solved exactly using a ME ansatz. 

\subsection{Numerical example}
\begin{figure}[h]
	\centering
	\includegraphics[width=\textwidth]{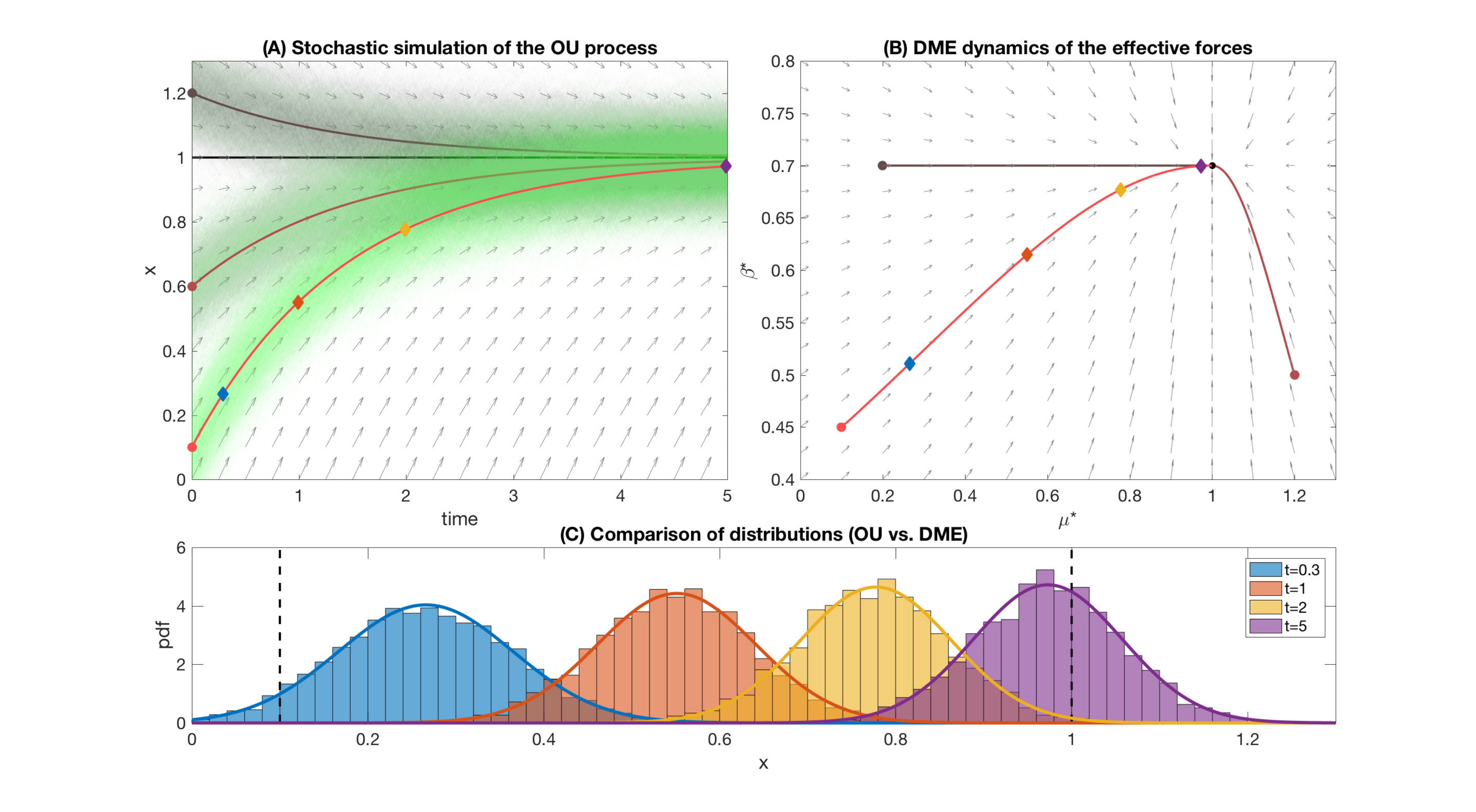}
	\caption{(A) Numerical simulations of OU process with parameters $\beta = 0.7$, $\mu = 1$, $\sigma = 0.1$.
	We used three random initial conditions from a distribution $\mathcal{N}(x_0,\sigma)$ with $\mu_0 = x_0 = 1.2, 0.6, 0.1$, 
	$\beta_0 = 0.7, 0.5, 0.45$ and $\sigma_0 = \sigma$.
	(B) Effective forces $(\mu^\ast,\beta^\ast)$ following dynamics \eqref{DME1}-\eqref{DME2} corresponding to 
	the same set of initial conditions as in panel A. 
	(C) Histograms of $x(t)$ at times $t=0.3,1,2,5$ (initial condition $\beta_0 = 0.45$, $\mu_0 = 0.1$, $\sigma_0 = 0.1$ as in panel A) 
	from the simulated data and approximated distributions \eqref{OU_stationary}
	for the effective forces. The time points correspond to the diamonds of matching color in the panels A-B. 
	Code in the Electronic Supplementary Information.
	\label{figOU}}
\end{figure}

Figure~\ref{figOU}A shows a numerical simulation of the OU process for three choices of initial Gaussian distribution (centered at $x_0=0.1,\;0.6,\;1.2$) for 3000 trajectories for each case (all parameters summarized in the figure legend). Initially, the system is in a stationary state corresponding to parameters $\mu_0\,\; \beta_0\,\; \sigma$ (Gaussian form with parameters $x_0,\; \sigma^2/2\beta_0$). However, at $t=0$ the parameters of the OU process rapidly change, pushing the system out-of-equilibrium. As a response, the distribution of sample trajectories follows a Gaussian form at each time, eventually converging to $\mathcal{N}(\mu,\sigma^2/2\beta)$. In figure~\ref{figOU}B we plot the 2-dimensional DME dynamics of effective forces $\beta^\ast$, $\mu^\ast$, which is exact. Each trajectory (vector field of the dynamical system is plotted as well) represents the complete solution of the FPE for a given parameter choice (at each time it is a Gaussian). The microscopic distribution at four different times in panel C shows an agreement between stochastic simulations (histograms) and microscopic distributions obtained from the DME approach (solid curves). In general, the goal of DME is to approximate the dynamics on the macroscale, thus, we do not expect DME to capture also the microscopic distribution. However, for the OU process DME is exact and thus the method recovers both the macroscale and the microscale properties of the process without loss of precision.

\section{Dynamics of a single island with immigration}
Here we use the DME method to approximate stochastic population dynamics. We will study a simple, yet nonlinear model -- stochastic logistic population growth in a single island with immigration from other habitats.

\subsection{The model}
In the case of unlimited resources and absence of predation, populations would grow indefinitely. However, in natural populations this is not the case: various factors impose bounds on this exponential growth. 
The logistic growth model describes population size regulation in the absence of demographic stochasticity. At low population sizes, when resources are abundant and competition is low, the population grows with its intrinsic growth rate, $r$.  However, the total growth rate of the population decreases linearly with increasing population size. In particular, the growth rate is zero when the population is at carrying capacity, $K$, reflecting the situation when each individual replaces itself in each generation. The carrying capacity represents the maximal sustainable population size. 
We follow the population in a single island, with a migration from other habitats at rate $m$. In the presence of demographic stochasticity the described population dynamics can be formulated using a stochastic differential equation
\begin{equation}
  \mathrm{d} n=\left[n(r-\lambda n)+m \right] \mathrm{d} t +\sqrt{\gamma n} \mathrm{d} \xi\,,
  \label{eq:logistic}
\end{equation}
where $n(t)$ represents the population size at time $t$,  $\lambda=r/K$ is the density regulation and $\gamma$ describes the variance in population size. For the sake of simplicity we fix $\gamma =1$ corresponding to a Poisson($1$) number of offspring for each individual (with total variance$n$). 
Extinction in this stochastic dynamics for $m=0$ is unavoidable from a mathematical point of view (as the process is a critical branching process) but can be prevented by migration when $m>0$.

In general, complex eco-evolutionary interaction requires including changes in population size due to demographic processes, changes occurring in gene frequencies due to selection, as well as the various feedback mechanisms connecting them. Feedback loops between population sizes and gene frequencies can result from migration and hard selection, i.e. that the size of the population depends on its 
genetic composition. Such questions were studied in \cite{szepetal2020}, but only in a stationary case. Relaxing the assumption of stationarity makes the problem more realistic, and inherently more difficult. The stochastic logistic dynamics with immigration \eqref{eq:logistic}, despite its simplicity, serves as the first step to understanding the biologically more realistic scenario of stochastic models in population dynamics \cite{Ewens2012}.
We are interested in how changes in the environment reflect on the dynamics of biological quantities, particularly when the system is out of equilibrium.
Since the model is nonlinear, the dynamics of moments, i.e., average population size, etc., are not closed. Nevertheless, the DME method can be applied to reduce the stochastic dynamics to a low-dimensional deterministic dynamics of the key observables.

Based on \eqref{eq:logistic}, we find the corresponding  FPE describing the time evolution of the probability 
distribution $u(t,n)$ (which is of the same form as \eqref{eq:FPE}):
\begin{equation}
\frac{\partial u}{\partial t} = - \frac{\partial}{\partial n} \left[ (n(r-\lambda n)+m)u\right] + \frac{1}{2} \frac{\partial^2}{\partial n^2} [n u].
  \label{eq:lg_FPE}
\end{equation}
The stationary solution of \eqref{eq:lg_FPE} can be found in the form of a potential function. Note that it indeed has the same form as the distribution that we observed earlier \eqref{eq:equilibrium} to maximize entropy:
\begin{equation}
u(n)=\frac{1}{\mathcal{Z}} \frac{1}{n}\mathrm{exp} \left\{ 2 (r n -\frac{\lambda n^2}{2}+m \log(n)) \right\} = \frac{1}{\mathcal{Z}}v(n)\mathrm{e}^{2 \ba \cdot \mathbf{A}}, 
\label{eq:stationaryLOG}
\end{equation}
where $v(n)=\frac{1}{n}$ is the baseline distribution (the stationary solution without any forces acting on the system), $\mathbf A=(n,-\frac{n^2}{2}, \log(n))$ is a set of observables, and $\ba=(r,\lambda,m)$ is a set of the ecological forces driving the 
system. The potential function $\boldsymbol{\alpha} \cdot \mathbf{A}$ consists of the effects of growth, density regulation, and migration.
We assume that migration is strong $m>1/2$ so even though the function $v(n)$ is not integrable on $\Omega_X = (0,\infty)$, the function $u(n)$ is integrable.
The expectations of the observables have biologically meaningful interpretations, and can, in principle, be measured. In our case, $\langle n \rangle$ corresponds to the expected population size, $\langle n^2\rangle$ to the second moment of population size, and the third term, $\langle \log(n) \rangle$ is the logarithm of the geometric mean of the population size. The normalizing constant $\mathcal{Z}$, which is the function of the effective forces $\alpha$, plays an important role, as a generating function for quantities of interest \cite{barton2009}
\begin{equation}
  	\frac{\partial \log(\mathcal{Z})}{\partial (2 \alpha_j)}=\langle A_j(n) \rangle,  \qquad 
  	\frac{\partial^2 \log(\mathcal{Z})}{\partial (2 \alpha_i)^2} = \mathrm{Cov}(A_i(n),A_j(n)) = \mathbf C_{i,j}.
  	\label{form_cov}
\end{equation}
Given a set of forces $\ba$, the system evolves to a stationary distribution \eqref{eq:stationaryLOG} that maximizes entropy with constraints on the observables, where $2\boldsymbol{\alpha}$ serve as the Lagrange multipliers. We are interested in how the dynamics change when the set of forces changes in time, and in the most extreme case when the set of initial forces $\boldsymbol{\alpha}_0$ change rapidly to a new set of values $\boldsymbol{\alpha}_1$. The observables will evolve towards the new stationary state, which creates a path between $\boldsymbol{\alpha}_0$ and $\boldsymbol{\alpha}_1$ in the space of effective forces. 

Under the diffusion approximation we can derive ordinary differential equations (similarly to \eqref{eq:Amean_simp} for the changes in the mean of the observables $\mathbf A=(n,-\frac{n^2}{2}, \log(n))$ 
\begin{align}\label{eq:odes}
  \frac{\mathrm{d}}{\mathrm{d}t} \langle n \rangle   &=r \langle n \rangle -\lambda  \langle n^2 \rangle +m, \\
  \frac{\mathrm{d}}{\mathrm{d}t}      \langle n^2 \rangle &= (2m+1)  \langle n \rangle +2r   \langle n^2 \rangle  
      -2 \lambda   \langle n^3 \rangle, \\
  \frac{\mathrm{d}}{\mathrm{d}t}        \langle \log (n) \rangle &= r- \lambda   \langle n \rangle +\left(m-\frac{1}{2}\right)  \left\langle \frac{1}{n} 
        \right\rangle,
\end{align}
where the choice of $\mathbf A$ follows from the stationary form \eqref{eq:stationaryLOG}.
This dynamical system is not closed, yet, we may apply the DME method to derive a 3-dimensional approximation for the dynamics of effective forces $\bast$ of the form \eqref{eq:DME} with a particular form of matrices $\mathbf {B}_{\bast}$ and $\mathbf {C}_{\bast}$ derived in Appendix \ref{ap:LG}. Note that the method is fully general and may be applied for arbitrary functions $\boldsymbol \alpha(t)$, capturing non-stationary ecological situations.

\subsection{Numerical example}

\begin{figure}[h!]
   \centering
\includegraphics[width=0.9\textwidth]{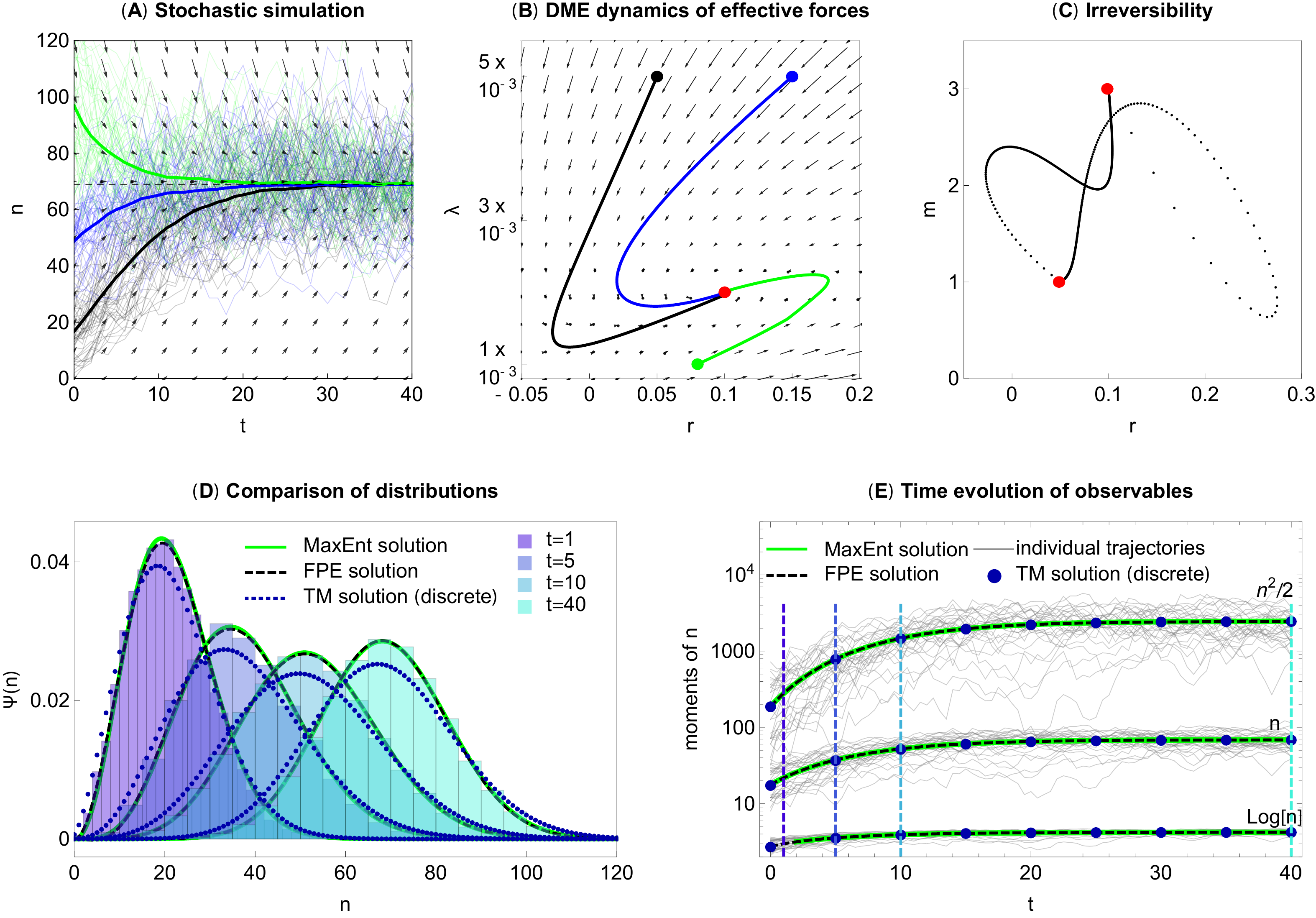}
\caption{ (A) Numerical simulations of stochastic population dynamics on a single 
island with immigration. Parameters are $\alpha_1=\{r,\lambda,m\}=\{0.1, 0.002, 3\}$. We 
used initial conditions, $\alpha_{0,1}=\{0.05, 0.005, 
1\}$ (black), $\alpha_{0,2}=\{0.15, 0.005, 5\}$ (blue), and $\alpha_{0,3}=\{0.08, 0.001, 2\} $
(green). (B) Corresponding dynamics of the effective forces projected to the $(r,\lambda)$ space.
(C) Irreversibility of the 
process: 2D projections of the trajectories between $\alpha_1=\{0.1, 0.002, 3\}$ and $\alpha_{0,1}=\{0.05, 0.005, 1\}$ 
and reversed are not the same. 
(D) Histograms of population sizes at 
$t=1,5,10,40$ with initial condition $\alpha_{0,1}=\{0.05, 0.005, 1\}$ (black curves in panels A-B). The numerical solution of the corresponding FPE, 
the discrete transition matrix prediction, and the DME all show a close match. (E) 
The three observables $n$, $\log(n)$, and $n^2/2$.
Code in the Electronic Supplementary Information.
 }
\label{fig:graph}
\end{figure}

To understand the relationship between the dynamics of the original system \eqref{eq:logistic} and the 
dynamics of the reduced system
we simulated individual 
population size trajectories using the Euler-Maruyama method and then compared them to 
the predictions of the DME method. In figure \ref{fig:graph}A, we see the simulated trajectories for three sets of 
initial conditions. The initial conditions were not fixed, instead, 
they were randomly drawn from a stationary initial distribution, parametrized by 
the growth rate ($r_0$), the strength of density regulation ($\lambda_0$), and migration ($m_0$). 
Starting in equilibrium, we changed the environmental forces abruptly 
to new values $\boldsymbol{\alpha} = (r,\lambda,m)$ at time $t=0$. This forced the system out of equilibrium 
and shifted the trajectories toward the new equilibrium, independent of the initial condition.

Instead of using the original stochastic differential equation, in the DME we follow the dynamics of the effective forces $\boldsymbol{\alpha}^\ast = (r^\ast,\lambda^\ast,m^\ast)$ as they move along to the new equilibrium
shown in figure \ref{fig:graph}B. Note, that the parameter space is 3 dimensional, but only a $2$ dimensional projection is presented 
here. A single point in this space describes the full distribution of population sizes.

Figure \ref{fig:graph}C shows that dynamics of the effective forces in the DME approximation of the stochastic logistic model with
migration is irreversible, i.e., the trajectory in the space of effective forces when the system changes from $\ba_0$ to $\ba_1$
is different from changing $\ba_1$ to $\ba_0$. In both cases the system was initialized with the stationary distribution and
the forces were changed at time $t=0$.

But how close is the distribution approximated by DME to the real distribution? 
Despite the simple form of our equations, it is not possible to solve it explicitly analytically. 
Thus, we compared the numerically computed distributions for the original (i.e., exact) problem 
with the numerically computed distributions obtained by the DME approximation (in figure \ref{fig:graph}D). 
We used three approximations to solve the original problem: 
(1) using the trajectories from figure \ref{fig:graph}A (approaching the exact distribution when the
 time step is small and the number of trajectories is large), 
(2) numerically solving the FPE for the process by the native solver of \textit{Mathematica}, and 
(3) using the transition matrix method, where we follow a Markov-chain, a continuous time  birth-death 
process on the discrete space of non-negative integers. 
On the other hand, we used an Euler scheme to solve the DME system.

Although all methods are approximate, the original problem can be solved 
with any precision using methods (1-2) and thus we may focus here on the accuracy of the DME method itself. 
We find that they are in a good agreement with each other, the only exception being the transition matrix 
method, which is defined on a discrete space (and to retain biological meaning also has a slightly different variance from \eqref{eq:logistic}). We compared the distributions at different time points in panel D. 
We observe that although the transition in the observable quantities is rather slow and monotonic, 
the changes in the effective forces can be abrupt and non-monotonic. 
Note, that the DME method does not guarantee 
that the microscopic population size distributions are identical, in fact, they can differ substantially. 
Nevertheless, the DME method aims to capture the agreements between the macroscopic variables. 
Figure \ref{fig:graph}  (E) shows all three key observables
and shows that the DME method is in an excellent agreement 
with the model.

\begin{figure}[h!]
   \centering
\includegraphics[width=\textwidth]{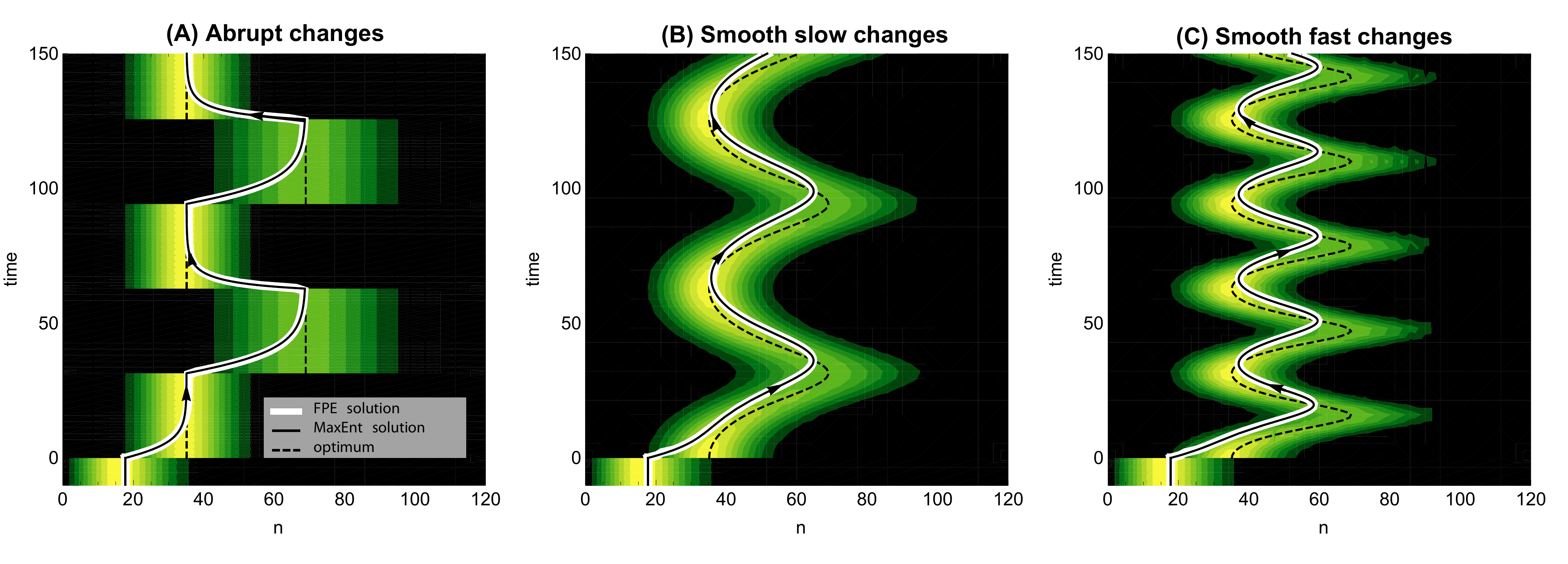}
\caption{Periodic changes in the carrying capacity between $20$ and $50$.
The system starts from equilibrium with parameters $\{0.05,0.005,1\}$ (as in Figure~\ref{fig:graph}), then periodic shifts
occur between $\{0.1,0.0005,3\}$ and $\{0.1,0.0002,3\}$. The 
equilibrium distribution of population size is shown as it changes in time (background colors). The 
black dashed line is the expectation of these distributions, the black solid 
line shows the solution of the DME, whereas the white is the solution of the 
FPE. 
 }
\label{fig:snakePlot}
\end{figure}

In the previous example we demonstrated that the method works well in the most extreme case, 
when the forces in the dynamics change abruptly. This is surprising, as the DME approximation 
is based on a quasi-stationary assumption, which would intuitively work only when the forces are changing 
slowly. This may suggest that the method will perform even better under slow environmental changes, 
which in reality may be more likely to happen than an abrupt change. Temporal differences in the environment can 
be abrupt, causing populations to become maladapted and possibly drive them to extinction. Less rapid environmental changes 
can be observed on various timescales, for example the warming of the oceans \citep{Munday2013} or the yearly cycle of 
seasons \citep{Kingsolver2017}, resulting in different migration rates throughout the year due to varying resource abundance.

In figure \ref{fig:snakePlot} we compare periodic changes for three scenarios: an abrupt, a slow and a fast but 
continuous change. We found again that the solution of DME is in a good agreement with that of the FPE, 
with the temporal dynamics in the DME and the moments in the FPE equation being indistinguishable by eye. 
We also see that the solution of the non-equilibrium dynamics lags behind the equilibrium 
of the environment, and that the amount of this lag depends on the speed of change of the ecological forces.
Moreover, faster environmental oscillations also result in a smaller range of effective forces. 
In the extreme case of very fast environmental changes 
(e.g. oscillations with large frequency, or environmental temporal noise of a fixed variation)
one expects that effective forces will stay almost stationary as the convergence to equilibrium is much slower than the timescale
of environmental fluctuations.

\section{Discussion}

We presented an application of a DME method, which helps reduce complexity of the stochastic process by linking the microscopic quantities to the macroscopic observables using a dynamic modification of the maximum entropy method. 

We first studied the OU process after a rapid change in its parameters. This example demonstrates our main strategy for understanding non-stationary dynamics -- instead of following the full stochastic dynamics to follow just the key observables (in this case the first two moments), which change deterministically. The observables and the forces acting on them were identified from the potential form of the stationary solution, which is also a solution of a maximum entropy problem and for the OU process has a Gaussian form. The dynamic problem was solved by the DME method that uses the stationary ansatz but allows the forces to change in time to best approximate the dynamics of observables. We derived a two-dimensional dynamical system of the effective forces that characterize the solution of the OU dynamics. Despite the intricacies of the DME approximation, the DME dynamics coincides with an exact solution of the OU process. This is because the dynamical equations for the first two moments are closed and thus the solution is Gaussian at all times. However, even though the OU dynamics are linear, the effective forces solve a nonlinear system of ordinary differential equations.

The focus of our work is on the stochastic island model, represented by nonlinear population dynamics based on a logistic equation supplemented by migration from other islands. The key parameters of the problem, the intrinsic growth rate, the carrying capacity, and the migration rate, are in general functions of time, reflecting temporal environmental changes, which make the problem out-of-equilibrium. Nonlinearity of the process results in the dynamics of moments $\langle n^k \rangle$ which are not closed for any k. Therefore we used DME to derive a 3-dimensional nonlinear dynamical system for the effective forces using DME. The associated observables are no longer just the first three moments but include $\langle n \rangle$, $\langle -n^2/2 \rangle$ and $\langle \log n \rangle$. Unlike for the OU model, the DME approximation of the island model with migration is no longer exact. The system is not fully explicit but contains terms using hypergeometric functions. Nevertheless, it can be solved for dynamic environmental forces using standard numerical solvers. 

We found that the effective forces in the DME approximation lag behind the true environmental forces, which is more pronounced when the environmental forces change faster. However, even in cases of rapid changes of the environmental forces the observables in the DME approximation are still extremely accurate at all times and thus the effective forces serve as a proxy for the dynamics. When the environmental forces settle to constant values, the effective forces also converge to these values. The DME serves as a change of optics where we represent the non-equilibrium dynamics using a series of equilibria parametrized by dynamical effective forces. The strength of the DME method in our view is not in speeding up the numerical method for solving the problem but in understanding the underlying dynamics in an appropriate low-dimensional space.

Although it is possible to use the quasi-stationary approximation without the connection to the ME, physics provides us with useful information. By formulating the suitable ME problem for the stationary distribution we learned which macroscopic quantities are important for the low-dimensional projection of the dynamics. Moreover, the observables and the environmental forces are dual variables in the appropriate ME formulation. While the observables enter the variational problem via the constraints, the forces are the corresponding Lagrange multipliers in ME. 

In addition, the DME method can be placed into the arsenal of methods for non-equilibrium dynamics, as we have shown in figure~\ref{fig:methods}. It is built from the stationary ME method but unlike the MC method, which is essentially a ME method applied on temporal trajectories, it uses the FPE to establish relationships between the time points.

One of the most striking properties of the DME method is its accuracy on the macroscopic level. This is surprising because the quasi-stationary assumption suggests validity of the approach when the applied forces change adiabatically. However, even for the fast changing forces the approximation stays very accurate. The unusual form of the DME approximation makes the analytical study of the accuracy of the method a difficult mathematical problem which remains an open problem to this date despite insight provided in \cite{Bodova2018}.

This work outlines the first step towards studying more complex questions where the complexity of the problem is prohibitive for studying the full problem. In particular, our future goal is to explore eco-evolutionary dynamics where the ecological and population genetic timescales interact. Such interaction has been studied in \cite{szepetal2020} but only in the stationary case. Although the existence of an explicit stationary distribution in principle allows us to explore the dynamics in the non-stationary environment using DME, the structure of the problem poses multiple difficulties that need to be resolved first. 

The approach may also be suited to stochastic problems in different disciplines. The method is based on the structure of the problem, in which the stochastic dynamics are described by the FPE and the stationary solution is explicit. This includes a wide range of problems accross disciplines, for example stochastic coagulation-fragmentation dynamics when the rates satisfy a detailed balance condition (existence of an explicit stationary distribution for this problem was shown in \cite{Durret2009}).




\section*{Funding} This work has been supported by the Scientific Grant Agency of the Slovak Republic under the Grants Nos. 1/0755/19 and 1/0521/20.


\appendix

\renewcommand{\thesection}{\Alph{section}}
\numberwithin{equation}{section}

\section{DynMaxEnt for the OU process, derivation of the $\mathbf B$ and $\mathbf C$ matrices}\label{ap:OU}

We briefly outline the key steps in the derivation of DME. The forces and observables are $\ba = (2\mu\beta,-\beta)$ and $\mathbf A = (x,x^2)$. The expectations $\langle A_i \rangle$ follow a closed system of ODEs 
\begin{equation}
	\langle \mathbf A \rangle'= 
	\begin{pmatrix}
	1/2 & \langle x \rangle \\
	 \langle x \rangle &  2\langle x^2 \rangle
	\end{pmatrix}
	\boldsymbol \alpha +
	\begin{pmatrix}
	0 \\
	\sigma^2
	\end{pmatrix} = \mathbf B \boldsymbol \alpha + \mathbf V \,.
	\label{Beqn}
\end{equation}
To apply the DME approximation we assume that at every time there are effective forces $\bast$ such that $\mathbf B_{\bast} {\bast}+ \mathbf V^\ast = 0$ (with the moments in matrix $\mathbf B$ evaluated at the stationary distribution $\bar u_{\bast}$). Using stationarity condition $\mathbf V^\ast = -\mathbf B_{\bast} {\bast}$ to substitute $\mathbf V$ by $\mathbf V^\ast$ and $\mathbf B$ by $\mathbf B_{\bast}$ in \eqref{Beqn} we obtain
\begin{equation}
	\frac{\mathrm{d}}{\mathrm{d}t} \langle \mathbf A \rangle= 
	\mathbf  B_{\bast}  (\boldsymbol \alpha -\bast) \label{Aeqn}\,.
\end{equation}
The matrix $\mathbf  B_{\bast} $ can be expressed in terms of the effective forces $\bast$, although we will write most of the expressions in terms of $\mu^\ast$ and $\beta^\ast$ (the transformation from $\bast$ to $(\mu^\ast,\beta^\ast)$ is regular). The matrix $\mathbf  B_{\bast} $ can be expressed as
\begin{equation}
	 \mathbf B_{\bast}=
	\begin{pmatrix}
	1/2 & \mu^\ast \\
	 \mu^\ast &  2(\mu^\ast)^2+\sigma^2/\beta^\ast
	\end{pmatrix}\,.
\end{equation}
Finally, we change variables in \eqref{Aeqn} using the scaled covariance matrix $\{\mathbf C_{\ba^\ast}\}_{ij} = \Cov(A_i,A_j)/\sigma^2$
to obtain dynamics of $\bast$
\begin{equation}
	\mathbf C_{\bast} = \frac{\mathrm{d}\bAst}{\mathrm{d}\bast} = 
	\frac{1}{2\beta}
	\begin{pmatrix}
	1 & 2\mu\\
	2\mu & \sigma^2/\beta + 4 \mu^2 
	\end{pmatrix}
	\,, \qquad
	\mathbf C_{\bast}^{-1} = \frac{2\beta}{\sigma^2}
	\begin{pmatrix}
	4\beta\mu^2 + \sigma^2 & -2\mu\beta\\
	-2\mu\beta & \beta
	\end{pmatrix} \,,
\end{equation}
where for simplicity we dropped the $^\ast$ notation in the above equations (all coefficients $\mu$ and $\beta$ are understood as $\mu^\ast$ and $\beta^\ast$). Plugging this into \eqref{Aeqn} (this time keeping all $^\ast$ symbols) leads to 
\begin{align}
	\frac{\mathrm{d}\bast}{\mathrm{d}t}&=\mathbf C_{\bast}^{-1} \mathbf B_{\bast} (\ba-\bast) = 2\beta^\ast
	\begin{pmatrix}
	\frac12 & -\mu^{\ast}\\
	0 & 1
	\end{pmatrix}
	\begin{pmatrix}
	2\beta\mu - 2\beta^\ast\mu^\ast \\
	\beta^\ast - \beta
	\end{pmatrix}\,.
\end{align}
After some algebraic manipulation the dynamics of $\bast$ becomes
\begin{align}
	\frac{\mathrm{d} }{\mathrm{d}t} (2\mu^\ast \beta^\ast)&= 2\beta^\ast (\mu\beta-\mu^\ast\beta^\ast) + 2\mu^\ast \beta^\ast (\beta-\beta^\ast) \,, \label{DMEfull1}\\
	\frac{\mathrm{d} }{\mathrm{d}t} (-\beta^\ast) &=  2\beta^\ast (\beta^\ast - \beta)\,, \label{DMEfull2}
\end{align}
Finally we transform the dynamics of $\bast$ to $\mu^\ast$, $\beta^\ast$
\begin{align}
	\frac{\mathrm{d}\mu^\ast}{\mathrm{d}t}&= \beta (\mu-\mu^\ast) \,,\\
	\frac{\mathrm{d}\beta^\ast}{\mathrm{d}t}&=  2\beta^\ast (\beta - \beta^\ast)\,.
\end{align}
This system is identical to \eqref{DME1}-\eqref{DME2}.

\section{DynMaxEnt for the island model, derivation of the $\mathbf B$ and $\mathbf C$ matrices}\label{ap:LG}

Equation \ref{eq:odes} can be written using the matrix notation:
\begin{align}
  \label{matEq}
	\langle \mathbf A \rangle'&= 
	\begin{pmatrix}
	\frac{1}{2} \langle n \rangle & \langle n^2 \rangle & \frac{1}{2}\\
	 \langle n^2 \rangle &  2\langle n^3 \rangle & \langle n \rangle \\
	 \frac{1}{2}  &  \langle n \rangle & \frac{1}{2} \left\langle \frac{1}{n} \right\rangle \\
	\end{pmatrix}
	\boldsymbol \alpha +
	\begin{pmatrix}
	0 \\
	 \langle n \rangle \\
           -\frac{1}{2}\left\langle \frac{1}{n} \right\rangle
	\end{pmatrix} = \mathbf B \boldsymbol \alpha + \mathbf V. 
\end{align}
At each time point, we approximate the elements of $\mathbf B$ and $\mathbf V$ using the stationary approximation $\mathbf B_{\bast} {\bast}+ \mathbf V^* = 0$. 
Substituting $\mathbf V = - \mathbf B_{\bast} {\bast} $ into (\ref{matEq}) and using that $\mathbf B \approx \mathbf B_{\bast}$, we obtain
$\frac{\partial    \langle A_i(n) \rangle}{\partial t} \approx \sum_j B^*_{i,j}   (\alpha_j-\alpha^*_j) $. Change of variables yields
\begin{equation}
 \frac{\dint \bast}{\dint t}  =\left[ \frac{d\bAst}{d\bast} \right]^{-1}   \frac{ \dint\bAst}{\dint  t}.
\end{equation}
The expectations of various functions of variable $n$ appearing in matrices $\mathbf  B_{\bast}$ and $\mathbf{C_{\bast}}$ can be expressed analytically under the condition that the migration rate is not too low ($m>\frac{1}{2}$).  Let us call the $\mathrm{k}^{th}$ moment of the stationary distribution $G(k)$, this can be expressed analytically in terms of hypergeometric functions (suppressing $^\ast$ notation)
\begin{align}
  \label{mom}
G(k ) &=\int_0^\infty n^k v(n)\mathrm{e}^{2 \ba \mathbf{A}} = \int_0^\infty n^{k+2m-1}\exp\{-\lambda n^2+2rn\} = \frac{1}{2} \lambda^{\frac{1}{2}(-1-k-2m)}\times \\
&\times \left( \sqrt{\lambda} \Gamma \left(\frac{k}{2}+m \right)
\, _1F_1 \left(\frac{k}{2}+m,\frac{1}{2},\frac{r^2}{\lambda}\right) \right. 
+ \left. 2 r \Gamma
  \left(\frac{k+1}{2}+m\right) \, _1F_1\left(\frac{k+1}{2}+m;\frac{3}{2};\frac{r^2}{\lambda 
  }\right)\right),
 \end{align}
 if  $\mathrm{Re}( k+ 2m / \gamma) >0$. Using the function $G$, all the moments of 
interest can be expressed as
\begin{equation}
  	G(0)=\mathcal{Z}, \qquad 
  	\frac{G(1)}{G(0)}=\langle n \rangle \,,  \qquad 
  	\frac{G(2)}{G(0)}=\langle n^2 \rangle \,, \qquad
  	\frac{G(3)}{G(0)}=\langle n^3 \rangle\,, \qquad
  	\frac{G(-1)}{G(0)}=\left\langle \frac{1}{n} \right\rangle\,.
\end{equation}
Thus
\begin{align}
  	\label{matBlog}
	\mathbf B_{\bast}&= 
	\frac{1}{G(0)} \begin{pmatrix}
	G(1)& -G(2) & 1\\
	 -G(2)&  G(3) &G(1) \\
	1  &  -G(1) & G(-1) \\
	\end{pmatrix}\,.
\end{align}
Furthermore, we can express $\langle \log(n) \rangle$ analytically by taking the $j^{th}$
derivative of $G(k)$ with respect to $m$:
 \begin{equation}
  H(k,j)=\bbbe(n^k \log(n)^j)=\frac{1}{2^j} \frac{\partial^{(j)}G_k}{\partial 
  m^j}.
\end{equation}
The covariance matrix of the observables evaluated at the quasi-stationary distribution parametrized by the effective forces can then be written as
\begin{equation}
	\label{matClog}
 	\mathbf{C_{\bast}}= \left(
	\begin{array}{ccc}
 	\frac{G(2)}{G(0)}-\frac{G(1)^2}{G(0)^2} & \frac{1}{2} \left(\frac{G(1) G(2)}{G(0)^2}-\frac{G(3)}{G(0)}\right) 
 	& \frac{H(1,1)}{H(0,0)}-\frac{G(1) H(0,1)}{G(0) H(0,0)} \\
 	\frac{1}{2} \left(\frac{G(1) G(2)}{G(0)^2}-\frac{G(3)}{G(0)}\right) 
 	& \frac{1}{4} \left(\frac{G(4)}{G(0)}-\frac{G(2)^2}{G(0)^2}\right) & \frac{1}{2}
  	 \left(\frac{G(2) H(0,1)}{G(0) H(0,0)}-\frac{H(2,1)}{H(0,0)}\right) \\
 	\frac{H(1,1)}{H(0,0)}-\frac{G(1) H(0,1)}{G(0) H(0,0)} 
 	& \frac{1}{2} \left(\frac{G(2) H(0,1)}{G(0) H(0,0)}-\frac{H(2,1)}{H(0,0)}\right) &
  	\frac{H(0,2)}{H(0,0)}-\frac{H(0,1)^2}{H(0,0)^2} \\
\end{array}
\right)\,.
\end{equation}


\end{document}